\documentclass[AMA,Times1Col]{USG} 

\usepackage{algorithm}
\usepackage{amsmath}
\usepackage{cleveref}
\usepackage{graphicx}
\usepackage{bm}
\usepackage{multicol}
\usepackage{ulem}
\usepackage[utf8]{inputenc}
\usepackage{newunicodechar}
\newunicodechar{∼}{\textasciitilde{}}

\articletype{Original Article}

\received{Date Month Year}
\revised{Date Month Year}
\accepted{Date Month Year}
\journal{Journal}
\volume{00}
\copyyear{2026}
\startpage{1}

\begin{document}

\title{Modeling of plasma transport during edge-localized mode in tokamak using a kinetic Vlasov-Poisson code}

\author{Ce Wang}

\author{Sven Van Loo}

\author{Geert Verdoolaege}

\authormark{WANG \textsc{et al.}}
\titlemark{Modeling of plasma transport during edge-localized Mode in tokamak using a kinetic Vlasov-Poisson code}

\address{\orgdiv{Department of Applied Physics}, \orgname{Ghent University}, \orgaddress{\state{Ghent}, \country{Belgium}}}

\corres{Ce Wang, Department of Applied Physics, Ghent University, Ghent 9000 ,Belgium\email{Ce.Wang@UGent.be}}


\fundingInfo{China Scholarship Council (CSC) affiliated with the Ministry of Education of the P.R. China (File No. 202406060067) and the BOF of Ghent University (01SC4024)}

\abstract[Abstract]{A kinetic parallel transport code KOBRA based on a finite-volume method is developed to study edge localized mode (ELM) plasma transport from the mid-plane to divertor targets. The large scale separation between the Debye length (∼cm) and the connection length (∼10–20 m) leads to prohibitive computational cost in full 6D simulations. To alleviate this, an adaptive-mesh refinement (AMR) strategy is employed. Comparisons with uniform-grid simulations show that AMR accurately reproduces the characteristic ELM dynamics, including the rapid rise and slow decay of divertor fluxes, as well as the early-time peak induced by fast electrons. Analysis of the electron distribution and self-consistent electric field reveals that AMR efficiency is closely linked to phase-space evolution. Overall, AMR achieves comparable physical accuracy while reducing memory usage by $30\%-40\%$ and accelerating computations by up to a factor of two, demonstrating its effectiveness for high-dimensional kinetic ELM simulations.}

\articledoi{10.1002/ctpp.70156}
\maketitle



\section{Introduction}
\label{sec:Introduction}

The transient energy deposition on divertor targets caused by edge-localized modes (ELMs) represents one of the major challenges for the safe operation of tokamaks. On the one hand, ELMs induce intense heat loads onto the divertor target plates, which may cause melting and thereby shorten the divertor lifetime\cite{gunn2017surface}\cite{perillo2023measurements}. On the other hand, impurities generated by divertor erosion due to ELMs can be transported upstream, poisoning the core plasma and potentially leading to reaction \cite{borodkina2020estimation}\cite{putterich2008modelling}. Tungsten (W) has been widely selected as the primary divertor material in large-scale devices such as ITER\cite{hawryluk2009principal}, owing to its high melting point and low sputtering yield. Therefore, extensive experimental and simulation studies on tungsten wall damage induced by ELMs have been carried out in major tungsten-wall devices worldwide\cite{xu2021interpretive}\cite{hakola2021gross}. To achieve a deeper understanding of ELM experiments, physical models are required to investigate the transport of ELM plasmas and the consequent tungsten erosion. Various codes are available to simulate ELM transport and to identify the mechanisms underlying instability formation, such as the fluid codes BOUT++\cite{li2022simulation} and JOREK\cite{cathey2022mhd}. However, during ELMs, the pedestal region is characterized by high temperatures and low particle collisionality, suggesting that ELMs can be approximated as collisionless or weakly collisional plasmas. This provides evidence for the kinetic nature of ELM transport, while also suggesting that fluid codes may not be directly applicable. To date, there is a range of kinetic codes for ELM transport, including the PIC code BIT1\cite{tskhakaya2008self} and the kinetic codes VPM\cite{wang2024modeling} and VESPA\cite{manfredi2010vlasov}, which simulate the transport of ELM filaments in one dimension using a uniform grid. As the domain length is comparable to the connection length (which is of the order of 10-20 m in ITER) and much larger than the Debye length, $\lambda_D$, of the order of cm, this leads to a large computational cost in both time and memory, especially when modeling the whole ELM process in the full 6D case. Adaptive-mesh refinement (AMR) is able to address this as it allows for increased resolution in
regions where the distribution function exhibits sharp variations, enhancing numerical stability, while maintaining the original
resolution in smooth regions, thereby saving computational resources. 

We employ our recently developed kinetic code KOBRA based on the Vlasov equation to simulate plasma parallel transport along magnetic field lines during ELM and make the comparison between uniform grid and AMR grid on accuracy, memory use and computational speed. The organization of this paper is as follows: \Cref{sec:model} describes the KOBRA kinetic code. In \cref{sec:simulation}, KOBRA simulation results are compared with previous numerical studies. Finally, conclusions are presented in \cref{sec:Conclusion}.

\phantom{However, to date, studies employing kinetic approaches to investigate ELM transport in the SOL and pedestal regions remain very limited. Therefore, the development of a kinetic code to study the safe release of energy during ELMs and the physics associated with edge instabilities is essential for the high-performance operation of ITER.}

\section{\uppercase{simulation model}}
\label{sec:model}
\subsection{KOBRA}

KOBRA is a kinetic code based on the finite-volume method and solves the Vlasov–Poisson equations:

\begin{equation}
    \frac{\partial f_j}{\partial t}+v_x\frac{\partial f_j}{\partial x}-\frac{q_j}{m_j} \frac{\partial \phi}{\partial x}\frac{\partial f_j}{\partial v_x} =0,
    \label{eq:Vlasov}
\end{equation}

\begin{equation}
\frac{\partial^2\phi}{\partial x^2}=-\frac{1}{\varepsilon_0}\sum_j{q_j\int_{-\infty}^{+\infty}{dv f_j}},
    \label{eq:Poisson}
\end{equation}
here given in 1D1V and where $f_j$ is distribution function of particle species $j$, $q_j$ the particle charge, $m_j$ the particle mass and $\phi$ the electrostatic potential. Unlike particle-tracking Particle-In-Cell (PIC) codes, KOBRA directly processes the particle distribution function, enabling it to capture kinetic effects in fine phase-space structures while avoiding the numerical noise inherent in PIC simulations\cite{manfredi2010vlasov}. KOBRA includes cell-by-cell AMR\cite{AMR} where refinement is based on the variation of the distribution function. When the first two truncation errors of the distribution function on each cell exceed a specified tolerance, the cell is refined\cite{hittinger2013block}. 
Currently, KOBRA achieves second-order numerical accuracy in both phase space and time. 

While the above describes the default version of KOBRA, a slightly different approach is used when the numerical domain is much larger than the Debye length (as for the model in this paper). In this quasineutral case, the Poisson equation becomes singular and we then adopt the reformulated Poisson equation of Manfredi et al.\cite{manfredi2010vlasov} as well as their Asymptotic-Preserving (AP) scheme The AP scheme uses Strang splitting with a semi-Lagrangian advection scheme (although only 2nd order in KOBRA). This approach relaxes the stringent resolution requirements associated with the Debye length and electron plasma frequency\cite{degond2010asymptotic}. Although larger timesteps are allowed, we restrict the timestep so that the distribution function moves at most one grid cell. This is to minimise the regridding operations for the AMR, i.e. once after a full timestep, and to simplify the use for parallel computing.

\begin{figure}[!ht]
    \centering
    \includegraphics[width=\textwidth]{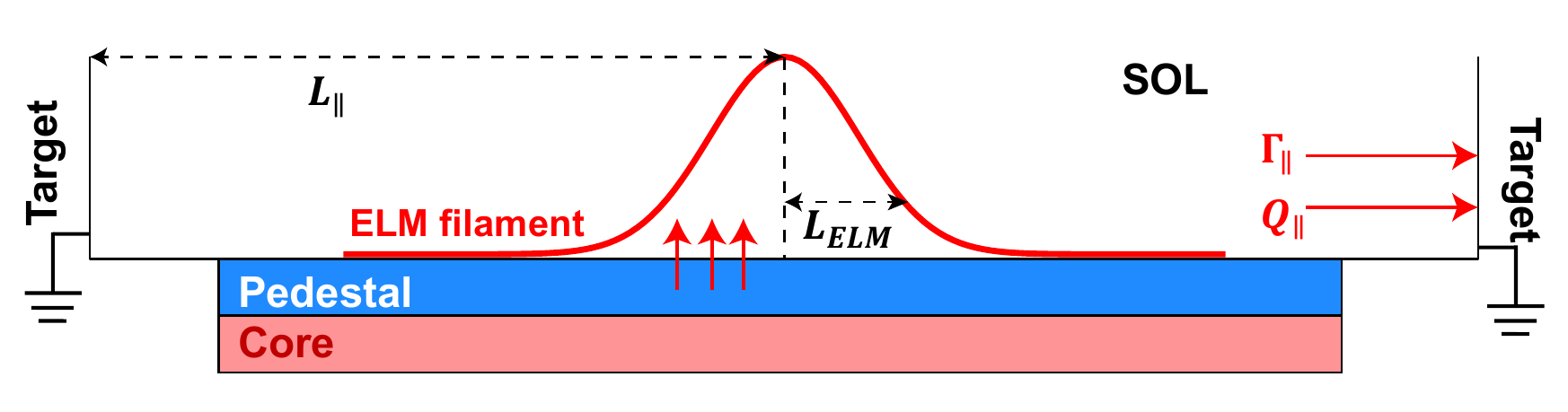}  
    \caption{Illustration of the spatial evolution of an ELM filament.}
    \label{fig:model} 
\end{figure}

\subsection{\uppercase{initial condition}}
We use the KOBRA to simulate the one-dimensional evolution of an ELM with length $L_{ELM}$, entering the SOL instantaneously from the pedestal region and transporting along magnetic field lines toward both divertor targets as illustrated by \Cref{fig:model}.
In the simulation, the $x$-direction represents the parallel direction, while particles in the perpendicular direction are assumed to always follow a Maxwellian distribution. Therefore, the distribution of charged particles in four-dimensional phase space $(x, \bf{v})$ can be expressed as $f_j(x,\textbf{v},t)=F_j(x,v_x,t)g_j(v_\perp)$, where $\displaystyle g_j(v_\perp)=\frac{m_j}{2\pi T_{ped}}\exp\left(-\frac{m_jv_{\perp}^{2}}{2T_{ped}}\right)$ with $j$ denoting either electrons or ions. Here, $T_{ped}$ is the electron temperature at the top of the pedestal, and the ion pedestal temperature is assumed to be equal to $T_{ped}$. Based on the above assumptions, the evolution of the particle distribution $F_j (x,v_x,t)$ in the parallel direction is governed by Eq. \ref{eq:Vlasov} and  the reformulated version of Eq.\ref{eq:Poisson}.

The instantaneous ELM burst from the mid-plane is described by the initial distribution $F_j(x,v_x,0)$
\begin{equation}
F_j (x,v,0)=\frac{n_{ped}}{\sqrt{2\pi}v_{Tj}} \exp\left(-\frac{x^2}{2L_{ELM}^2}\right)\exp\left(-\frac{v_x^2}{2v_{Tj}^2}\right).
    \label{eq:source}
\end{equation}
The spatial distribution of $F_j(x,v_x,0)$ is a Gaussian profile with a width of $L_{ELM}= 0.1L_{\parallel}$, where $L_{\parallel}$ is the parallel connection length. $n_{ped}$ denotes the electron density at the top of the pedestal, with the ion pedestal density assumed equal to $n_{ped}$, while the thermal velocity of the charged particles is given by $v_{Tj}=\sqrt{\frac{kT_{ped}}{m_j}}$. This numerical setup follows the VESPA configuration for type-I ELM transport on JET which correspond to background parameters of $T_{ped}=1.5\mathrm{keV}$, $n_{ped}=5\times10^{19} \mathrm{m^{-3}}$, and $L_{\parallel}=30\mathrm{m}$. As ion species we use ionised hydrogen ($m_{H^+}=1836m_{e^-}$).

The numerical domain is set to $x\in{[-L_{\parallel}, +L_{\parallel}]}$ and $v\in [-8v_{Tj}, 8v_{Tj}]$ for ions and electrons separately. Absorbing wall boundary conditions are imposed for $x$, meaning that there is no particle influx on both targets, and free-flow for $v$. The potentials at both divertor walls are grounded, i.e. $\phi = 0$. In order to set the model resolution, it is worth noting that a realistic value of $\lambda_D/L_{\parallel} \approx 10^{-6}$ is computationally restrictive when simultaneously resolving $\lambda_D$. Moulton et al. verified the impact of different ratios, i.e. $10^{-3}$ down to $10^{-6}$, on the particle and heat flux, defined by 
$\displaystyle \Gamma_{j} = \iiint f(\pm L_{\parallel}, \bm{v}, t)\, v_x \, d\bm{v}$  
and
$\displaystyle Q_{j} = \iiint \tfrac{1}{2} m_j \left( v_x^2 + v_{\perp}^2 \right) 
f(\pm L_{\parallel}, \bm{v}, t)\, v_x \, d\bm{v}$,
on the divertor walls with both VESPA and BIT1 codes\cite{moulton2013quasineutral}. They find no significant difference in the fluxes in this range. Therefore, we also adapt $\lambda_D/L_{\parallel}=10^{-3}$ in our model. While this would require a minimum spatial resolution of $\sim 2000$ grid cells, and even more if the Debye length would be sufficiently refined, the AP scheme allows a lower resolution. We set the resolution in the uniform model for both $x$ and $v$ to 1024 meaning that $\Delta x \approx 2\lambda_D$. For the AMR model, we use a base grid of 64 grid cells in both $x$ and $v$ with 4 levels of refinement as to have the same effective refinement as the uniform model.

\section{Simulation results}
\label{sec:simulation}

In the following, we investigate the ELM transport validating the results of KOBRA with previously published results and compare the  uniform-grid and AMR-based simulations with each other.

\begin{figure}[!h]
    \centering
    \includegraphics[width=1.0\textwidth]{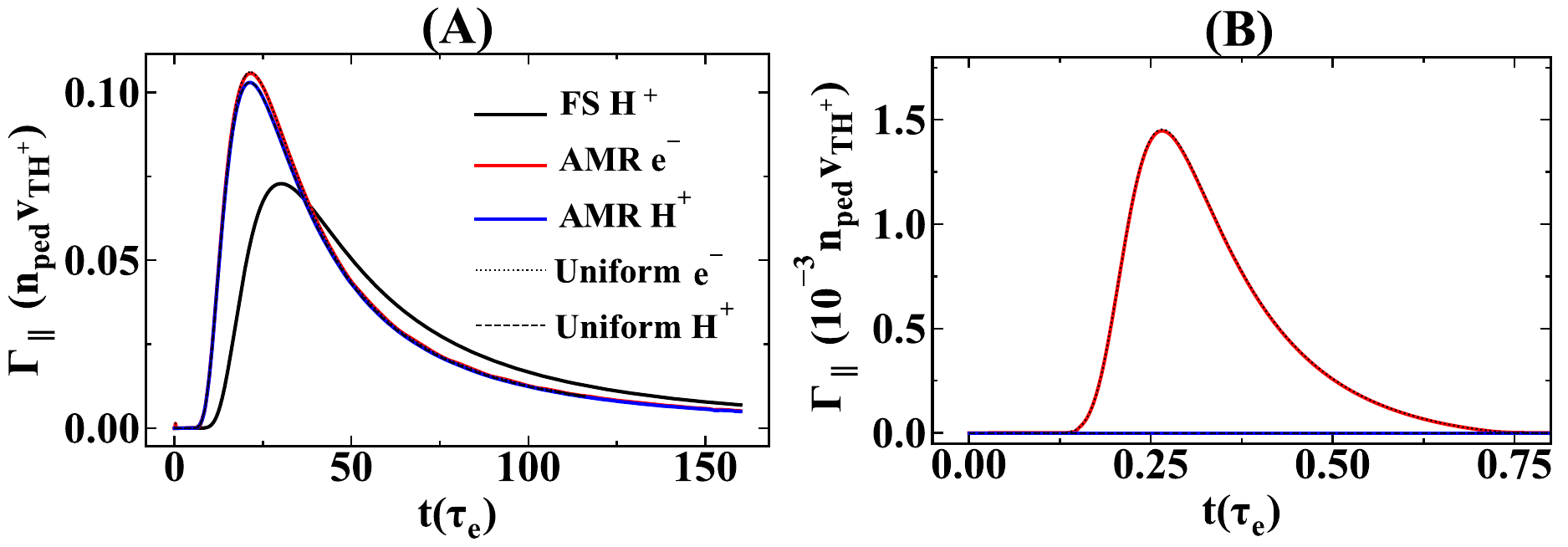} 
    \caption{(A) Parallel particle flux and (B) high-energy electron flux $\Gamma_{\parallel j}$ at the divertor target strike points for the uniform (dotted and dashed for electrons and ions, resp.) and AMR (red for electrons and blue for ions, resp.) simulations. The FS model shown in solid black. (B) Same as (A), but zoom in at the early evolution.}
    \label{fig:flux} 
\end{figure}

\begin{figure}[h]
    \centering
    \includegraphics[width=1.0\textwidth]{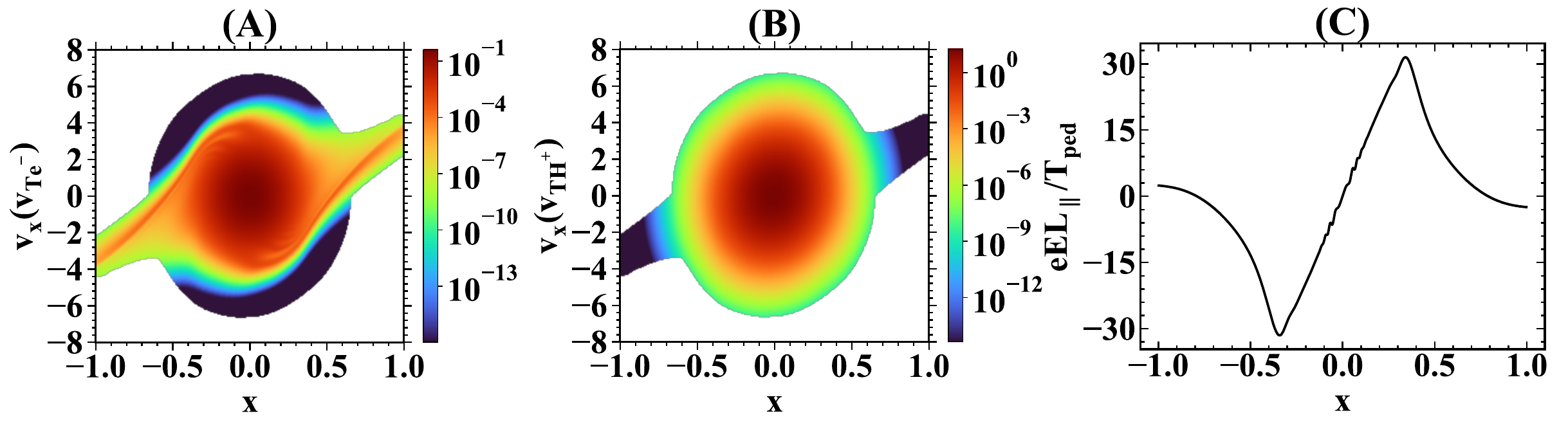}
    \caption{(A) Electron distribution function in phase space with logarithmic color scale and for the finest grid level, (B) the ion distribution (c) the corresponding electrostatic electric field in physical space at $t=0.20\tau_e$.}
    \label{fig:fast electron} 
\end{figure}

\begin{figure}[h]
    \centering
    \includegraphics[width=1.0\textwidth]{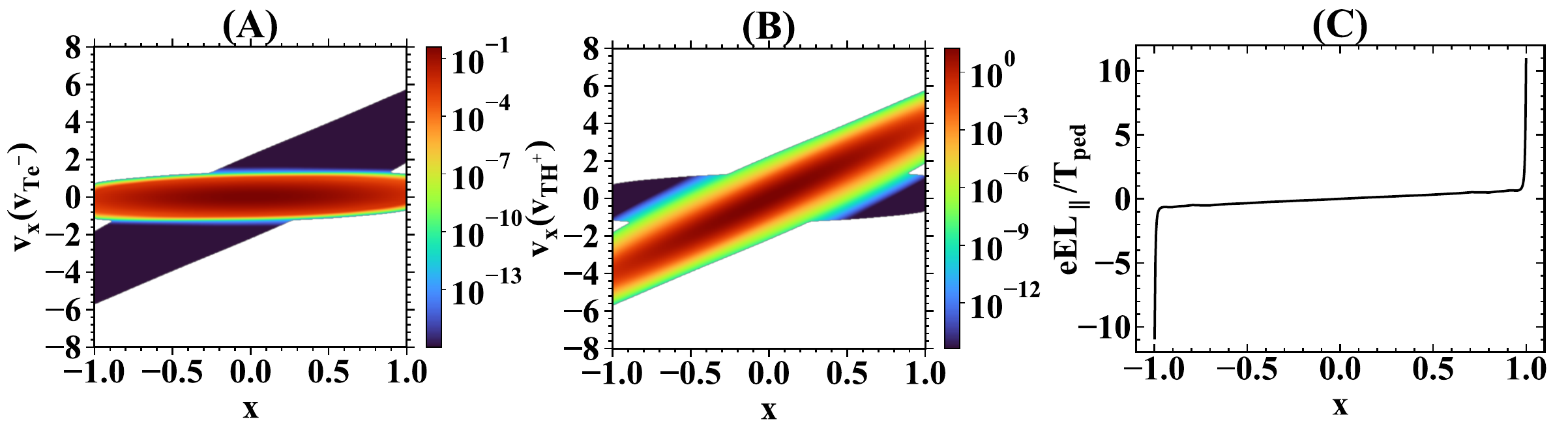}
    \caption{(A) Electron distribution function in phase space with logarithmic color scale and for the finest grid level, (B) the ion distribution (c) the corresponding electrostatic electric field in physical space at $t=10\tau_e$.}
    \label{fig:DistriAndEf} 
\end{figure}

Firstly, we model the ELM process to assess the capability of KOBRA in capturing the characteristic temporal behavior of ELMs, namely the rapid rise and subsequent slow decay of the divertor fluxes. \Cref{fig:flux} shows the parallel particle flux of ions and electrons at the strike points of the divertor target for the uniform and AMR simulations. The particle fluxes of electrons and ions are nearly identical reflecting the quasi-neutrality of the plasma. Only near the peak the electron flux is slightly higher. We can compare these flux rates with a free-streaming (FS) model. The FS model corresponds to a zero electric field/potential in the Vlasov equation). It yields a lower peak of $\Gamma_{ion^+}$ and at a later time. This difference arises as the ion acceleration by the self-consistent electric field is neglected in the FS model. This electric field is produced in the early phase of the ELM by a small population of fast electrons in the high-energy tail of the Maxwellian velocity distribution. These electrons can escape the ion confinement and reach the divertor wall before they can be captured. \Cref{fig:fast electron} clearly shows the stream of high-energy electrons (A) having reached the wall, while the ions (B) have not. Although the electron flux rate peak is only $10^{-3} n_{ped} v_{TH^+}$ (see \cref{fig:flux}B) it generates a charge separation and electric field (\cref{fig:fast electron}C). Note that this is a kinetic effect and cannot be reproduced by fluid models \cite{havlivckova2012comparison}. While the electric field initially accelerates the ions, it also decelerates the electrons, leading to a rapid compression of the electron distribution function in velocity space, i.e. after $10\tau_e$ the electrons have mainly speeds less than $v_{Te}$ (see \cref{fig:DistriAndEf}A). As the electron velocity decreases, most of the plasma in the domain becomes quasi-neutral, and consequently the electric field in physical space becomes weaker (\cref{fig:DistriAndEf}C). The later stages of the ELM transport can then be considered close to free-streaming for the ions. Only near the walls some charge separation is observed. As mentioned before, the overall charge density in the domain is positive and these positive charges form a localized sheath structure near the wall. We should note that these boundary layers restrict the use of larger spatial grid lengths. The boundary layer is of the order of a few $\lambda_D$ and a failure to resolve this produces unphysical oscillations near the wall.

\begin{figure}[h]
    \centering
    \includegraphics[width=1.0\textwidth]{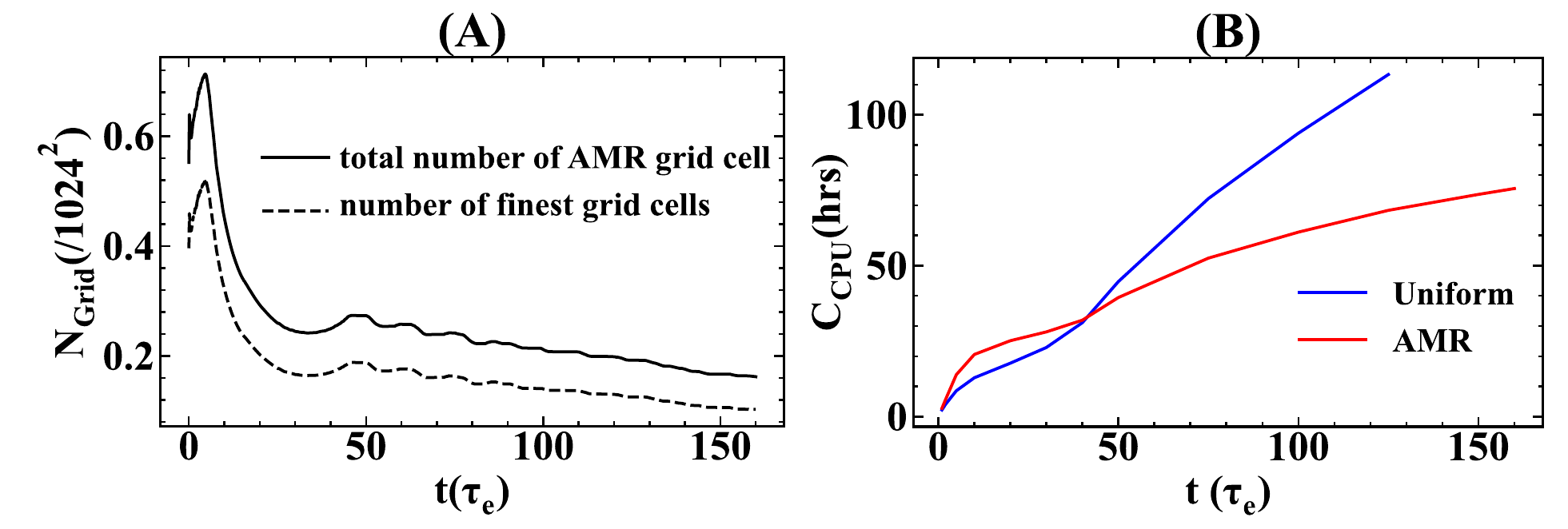} 
    \caption{Evolution of (A) the number of grid cells $N_{Grid}$ and (B) CPU time $C_{CPU}$ (single processor) as functions of the ELM duration for AMR and uniform grids.}
    \label{fig:Speed} 
\end{figure}

Our results match both qualitatively and quantitatively the results of Manfredi et al.\cite{manfredi2010vlasov} giving confidence that KOBRA can accurately model ELMs in the quasi-neutral regime. Furthermore, \cref{fig:flux} shows that the flux rates obtained for the uniform and AMR simulations are in close agreement, i.e. the difference is less than 1\%. This is because the finest grid covers the electron and ion distributions as can be seen in \cref{fig:fast electron} and \cref{fig:DistriAndEf}. However, there is a significant difference in computational performance.  \Cref{fig:Speed} shows the evolution of the number of grid cells $N_{Grid}$ and CPU time $C_{CPU}$ (for a single processor), respectively, for both AMR and uniform grids. For the uniform grid, $C_{CPU}$ increases roughly linearly with time as it has a constant number of grid cells. Variations in time are related to the computational time to solve the (reformulated) Poisson equation. 
In contrast, the AMR simulation is significantly slower initially and is about twice as slow as the uniform simulation at $10\tau_e$, but then starts to speed. At $t=40\tau_e$, the AMR and uniform simulation need the same CPU time and, as the simulation progresses to $120\tau_e$, the AMR becomes twice as fast as the uniform grid, i.e. the AMR takes 68hrs compared to 113hrs for the uniform model. Note that this is slower than the Manfredi et al. simulations as our timestep is more than an order of magnitude smaller than their fixed timestep of $\Delta t = 4\omega_{pe}^{-1}$. This simulation shows both the advantage and disadvantage of AMR. The speed of an AMR code is not only determined by the number of grid cells, but also by its distribution across the different grid levels. When the finest grid levels are significantly refined the overheads associated with AMR, i.e. advection of the distribution function on each level, regridding of each level and defragmentation, take much of the computational cost.  By examining the evolution of the number of cells on the finest level in \cref{fig:Speed}(A), we indeed see that, in the first $10\tau_e$, the finest grid covers about 50\% of the numerical domain and drops to about 20\% near $t=40\tau_e$.  The reduction in fine grid cells correspond to the increase in computational speed. While computational speed is one aspect, grid-based kinetic models also require significant memory storage. Again the number of grid cells is proportional to the memory cost. At the time when there is maximum grid refinement the AMR model uses about 70\% of the memory needed for the uniform grid (as the total number of grid cells on all levels is still lower than for the uniform model), while this reduces to about 18\% at the end of the simulation. Thus, AMR also reduces the meomory storage requirements.

\section{Conclusion}
\label{sec:Conclusion}

In this work, a kinetic parallel transport code, KOBRA, based on adaptive mesh refinement (AMR) and solved using the finite-volume method has been developed to investigate plasma transport from the mid-plane to the divertor targets during an ELM event. A systematic comparison of ELM transport and computational performance between AMR and uniform grids has been carried out on KOBRA. The results show that KOBRA accurately reproduces the evolution of the parallel particle flux $\Gamma_{\parallel j}$ during the ELM. Using AMR, KOBRA captures the main characteristic of an ELM, including the rapid rise and subsequent slow decay of the fluxes. Moreover, KOBRA with AMR successfully resolves the high-energy electron population at the early stage of the ELM, demonstrating its capability to capture fine-scale phase-space structures.

Furthermore, the analysis of the electron distribution function and electric field evolution reveals that the performance of the AMR method is strongly correlated with the ELM plasma dynamics. The rapid variation of the electron distribution driven by strong self-consistent electric fields at early times leads to intensive mesh refinement and higher initial computational cost, whereas the subsequent rapid decrease of number of grid cells enables substantial efficiency gains. As a result, the AMR approach eventually surpasses the uniform grid in computational efficiency while maintaining accuracy.

Overall, the present work demonstrates that AMR provides an efficient and physically reliable framework for kinetic simulations of ELM transport, offering a promising tool for high-dimensional phase-space problems.
Several improvements for KOBRA are planned for future work. First, the spatial numerical accuracy (currently second order) will be increased to fourth order to decrease the resolution on AMR grid to save memory for higher dimensional (i.e. 1D3V or 3D3V) ELM transport. Second, a collisional operator will be introduced to study the effects of interparticle collisions on energy deposition. Finally, KOBRA will be extended to different geometries, such as cylindrical coordinates.

\bmsection*{Acknowledgments}
Ce Wang gratefully thanks the China Scholarship Council (CSC) affiliated with the Ministry of Education of the P.R. China (File No. 202406060067) and the BOF of Ghent University (01SC4024) for their funding. VisIt, a parallel data visualization tool developed at the Lawrence Livermore National Laboratory (LLNL), was used to generate plots\cite{HPV:VisIt}.

\bmsection*{Conflict of interest}
The authors have stated explicitly that there are no conflicts of interest in connection with this article.

\bmsection*{DATA AVAILABILITY STATEMENT}
The data that support the findings of this study are available from the corresponding author upon reasonable request.

\bibliography{PET2025cewang}

\begin{thebibliography}{10}
\providecommand \doibase [0]{http://dx.doi.org/}%

\bibitem{gunn2017surface}
Gunn JP, Carpentier-Chouchana S, Escourbiac F, et al. Surface heat loads on the ITER divertor vertical targets. {\it Nuclear Fusion.} 2017\string;57(4)\string:046025.

\bibitem{perillo2023measurements}
Perillo R, Boedo J, Lasnier C, et al. Measurements and modeling of type-I and type-II ELMs heat flux to the DIII-D divertor. {\it Nuclear fusion.} 2023\string;63(8)\string:086031.

\bibitem{borodkina2020estimation}
Borodkina I, Borodin D, Brezinsek S, et al. Estimation of ELM effects on Be and W erosion at JET-ILW. {\it Physica Scripta.} 2020\string;2020(T171)\string:014027.

\bibitem{putterich2008modelling}
P{\"u}tterich T, Neu R, Dux R, et al. Modelling of measured tungsten spectra from ASDEX Upgrade and predictions for ITER. {\it Plasma Physics and Controlled Fusion.} 2008\string;50(8)\string:085016.

\bibitem{hawryluk2009principal}
Hawryluk R, Campbell D, Janeschitz G, et al. Principal physics developments evaluated in the ITER design review. {\it Nuclear Fusion.} 2009\string;49(6)\string:065012.

\bibitem{xu2021interpretive}
Xu G, Ding R, Ding F, et al. An interpretive model for the double peaks of divertor tungsten erosion during type-I ELMs in EAST. {\it Nuclear Fusion.} 2021\string;61(8)\string:086011.

\bibitem{hakola2021gross}
Hakola A, Likonen J, Lahtinen A, et al. Gross and net erosion balance of plasma-facing materials in full-W tokamaks. {\it Nuclear fusion.} 2021\string;61(11)\string:116006.

\bibitem{li2022simulation}
Li Y, Xia T, Zou XL, et al. The simulation of ELMs mitigation by pedestal coherent mode in EAST using BOUT++. {\it Nuclear Fusion.} 2022\string;62(6)\string:066018.

\bibitem{cathey2022mhd}
Cathey A, Hoelzl M, Harrer G, et al. MHD simulations of small ELMs at low triangularity in ASDEX Upgrade. {\it Plasma Physics and Controlled Fusion.} 2022\string;64(5)\string:054011.

\bibitem{tskhakaya2008self}
Tskhakaya D, Kuhn S, Tomita Y, Matyash K, Schneider R, Taccogna F. Self-Consistent Simulations of the Plasma-Wall Transition Layer. {\it Contributions to Plasma Physics.} 2008\string;48(1-3)\string:121--125.

\bibitem{wang2024modeling}
Wang C, Sang C, Liu J, Zhang C, Wang D. Modeling of tungsten divertor target erosion induced by impurity during edge-localized modes by using a kinetic model. {\it Contributions to Plasma Physics.} 2024\string;64(7-8)\string:e202300131.

\bibitem{manfredi2010vlasov}
Manfredi G, Hirstoaga S, Devaux S. Vlasov modelling of parallel transport in a tokamak scrape-off layer. {\it Plasma Physics and Controlled Fusion.} 2010\string;53(1)\string:015012.

\bibitem{AMR}
{Wareing} CJ, {Pittard} JM, {Falle} SAEG, {Van Loo} S. {Magnetohydrodynamical simulation of the formation of clumps and filaments in quiescent diffuse medium by thermal instability}. {\it Monthly Notices of the Royal Astronomical Society.} 2016\string;459(2)\string:1803-1818.
\newblock \href {\doibase 10.1093/mnras/stw581} {doi: 10.1093/mnras/stw581}

\bibitem{hittinger2013block}
Hittinger JA, Banks JW. Block-structured adaptive mesh refinement algorithms for Vlasov simulation. {\it Journal of Computational Physics.} 2013\string;241\string:118--140.

\bibitem{degond2010asymptotic}
Degond P, Deluzet F, Navoret L, Sun AB, Vignal MH. Asymptotic-preserving particle-in-cell method for the Vlasov--Poisson system near quasineutrality. {\it Journal of Computational Physics.} 2010\string;229(16)\string:5630--5652.

\bibitem{moulton2013quasineutral}
Moulton D, Ghendrih P, Fundamenski W, Manfredi G, Tskhakaya D. Quasineutral plasma expansion into infinite vacuum as a model for parallel ELM transport. {\it Plasma Physics and Controlled Fusion.} 2013\string;55(8)\string:085003.

\bibitem{havlivckova2012comparison}
Havl{\'\i}{\v{c}}kov{\'a} E, Fundamenski W, Tskhakaya D, Manfredi G, Moulton D. Comparison of fluid and kinetic models of target energy fluxes during edge localized modes. {\it Plasma Physics and Controlled Fusion.} 2012\string;54(4)\string:045002.

\bibitem{HPV:VisIt}
Childs H, Brugger E, Whitlock B, et al. VisIt: An End-User Tool For Visualizing and Analyzing Very Large Data.  2012\string:357-372.
\newblock \href {\doibase 10.1201/b12985} {doi: 10.1201/b12985}

\end{thebibliography}

\end{document}